\documentclass[10pt,shortpaper,twoside,web] {ieeecolor}

\usepackage{generic}
\usepackage{cite}
\usepackage{amsmath,amssymb,amsfonts}
\usepackage{algorithmic}
\usepackage{subfig}
\usepackage{graphicx}
\usepackage{textcomp}
\usepackage{psfrag} 
\usepackage{cuted}
\usepackage{float}
\usepackage{graphicx} 
\usepackage{amsmath}  
\usepackage[fancythm,fancybb,morse,ieee]{jphmacros2e} 
\usepackage{array}
\usepackage{bbm}
\usepackage{mathrsfs}
\usepackage{epstopdf}
\DeclareMathAlphabet{\mathpzc}{OT1}{pzc}{m}{it}
\usepackage{graphicx}
\usepackage{amsmath}
\usepackage[fancythm,fancybb,morse,ieee]{jphmacros2e}
\usepackage{caption}
\usepackage{subcaption}
\usepackage{multirow}
\usepackage{csquotes}
\usepackage{url}
\newtheorem{thm}{Theorem}
\newtheorem{lem}{Lemma}

\newtheorem{cor}{Corollary}
\newtheorem{defn}{Definition}

\newtheorem{rem}{Remark}

\def\BibTeX{{\rm B\kern-.05em{\sc i\kern-.025em b}\kern-.08em
		T\kern-.1667em\lower.7ex\hbox{E}\kern-.125emX}}
\begin{document}
	\title{Integration of Prior Knowledge into Direct Learning for Safe Control of Linear Systems}
	\author{Amir Modares, Bahare Kiumarsi, and Hamidreza Modares
		\thanks{ }
		\thanks{A. Modares is with Sharif University, Tehran, Iran (e-mail: amir.modares.81@gmail.com). B. Kiumarsi and H. Modares are with Michigan State University, USA (emails: kiumarsi@msu.edu; modaresh@msu.edu) }}
	
\maketitle
\thispagestyle{empty} 
\begin{abstract} 
This paper integrates prior knowledge into direct learning of safe controllers for linear uncertain systems under disturbances. To this end, we characterize the set of all closed-loop systems that can be explained by available prior knowledge of the system model and the disturbances. We leverage matrix zonotopes for data-based characterization of closed-loop systems and show that the explainability of closed-loop systems by prior knowledge can be formalized by adding an equality conformity constraint to the matrix zonotope. We then leverage the resulting constrained matrix zonotope and design safe controllers that conform with both data and prior knowledge. This is achieved by ensuring the inclusion of a constrained zonotope of all possible next states in a $\lambda$-scaled level set of the safe set. We consider both polytope and zonotope safe sets and provide set inclusion conditions using linear programming. 
\end{abstract}

\IEEEpeerreviewmaketitle

\section{Introduction}

\IEEEPARstart{D}ata-driven control design is categorized into direct data-driven control and indirect data-driven control \cite{Data1}-\cite{DI1}. The former parameterizes the controller and directly learns its parameters to satisfy control specifications. The latter learns a set of system models that explain the data and then designs a robust controller accordingly. 

A recent popular direct data-driven approach \textit{characterizes the set of all closed-loop systems} using data \cite{Data2}-\cite{Data6}. This approach has the potential to avoid the suboptimality of indirect learning \cite{Data4}. Besides, direct learning can achieve a lower sample complexity than indirect learning \cite{Data5}. However, despite advantages of direct learning, it is challenging to incorporate available prior knowledge into its learning framework.  Leveraging prior knowledge of the system model can significantly improve the performance of direct learning and reduce its conservatism. This occurs by removing the set of closed-loop systems that cannot be explained by prior knowledge. Despite its vital importance, the integration of direct learning and indirect learning is surprisingly unsettled.

In contrast to direct learning, indirect learning using system identification can incorporate prior knowledge to refine the set of learned system models. For instance, set-membership identification \cite{ID1}-\cite{ID2} can incorporate prior knowledge to learning the system models. Zonotope-based system modeling and its integration with prior knowledge have also been recently presented in \cite{ID21}-\cite{ID5}. 

This paper integrates closed-loop learning (i.e., direct learning) with prior knowledge (i.e., system model knowledge obtained from system identification and/or prior physical information) for linear discrete-time systems under disturbances. To this end, we characterize the set of all closed-loop systems that can be explained by the prior knowledge available in terms of the system model parameters and the disturbance bounds. To incorporate this prior knowledge into closed-loop learning, we first represent a matrix zonotope for closed-loop systems conformed with data. We then provide equality conformity constraints under which the explainability of closed-loop system models by prior knowledge can be formalized. We then leverage the resulting constraint matrix zonotope presentation of closed-loop systems to characterize the set of all possible next states by a constrained zonotope. This characterization is then leveraged to ensure the inclusion of the constrained zonotope of all possible next states in a $\lambda$-scaled level set of the safe set (i.e., to ensure $\lambda$-contractivity). The set inclusion conditions are provided for cases where the safe set is characterized by a convex polytope and by a constrained zonotope. \vspace{6pt}  

\noindent \textbf{Notations and Definitions.} Throughout the paper, $\mathbb{R}^n$ denotes vectors of real numbers with $n$ elements, $\mathbb{R}^{n \times m}$ denotes a matrix of real numbers with $n$ rows and $m$ columns. Moreover, $Q \ge  0$ denotes that $Q$ is a non-negative matrix with all elements being positive real numbers. The symbol $I$ denotes the identity matrix of the appropriate dimension.  The symbol $\bar{\textbf{1}}$ represents the column vector of values of one.
For a vector $x=[x_1,...,x_n]$, the notation  $\|x\|=\max \{x_1,...,x_n\}$ is used as the infinity norm, and $x_i$ denotes its $i$-th element. The unit hypercube in $\mathbb{R}^n$ is denoted by $B_{\infty}$. For a matrix $X$, $X_i$ denotes its $i$-th row, and $\|X\|$ denotes its infinity norm.
For sets $Z,W \in \mathbb{R}^n$, their Minkowski sum is $Z \oplus W= \{z + w |z \in Z,w \in W \}$. For a matrix $G=[G_1,...,G_s] \in \mathbb{R}^{n \times ms}$ with $G_i \in \mathbb{R}^{n \times m}$, \, $i=1,...,s$, and a matrix $N \in \mathbb{R}^{m \times p}$, we define $G \circ N=[G_1 N,...,G_s N] \in \mathbb{R}^{n \times ps}$, and $Vec(G)=[Vec(G_{1}),...,Vec(G_{s})]$, with ${Vec}(G_i)$ as an operator that transforms the matrix $G_i$ into a column vector by vertically stacking the columns of the matrix. 
 \vspace{3pt}

\begin{defn}
 \textbf{(Convex Polytope)}  Given a matrix $H \in \mathbb{R}^{q \times n}$ and a vector $h \in \mathbb{R}^q$, a convex polytope $\cal{P}(H,h)$ is represented by \vspace{-10pt}
\begin{align} 
\cal{P} (H,h) = \{ x \in {\mathbb{R}^n}:H x  \le h\}.
\end{align}
\end{defn} \vspace{6pt}

\begin{defn} \textbf{(Zonotope)} \cite{ID3}
 Given a generator matrix $G \in \mathbb{R}^{n \times s}$ and a center $c \in \mathbb{R}^n$, a zonotope $\cal{Z}=\big<G,c\big>$  of dimension $n$ with $s$ generators is represented by \vspace{-3pt}
\begin{align} 
& \cal{Z}=\big<G,c\big> \\ \nonumber & \quad = \big\{x \in \mathbb{R}^n:x=G \, \zeta +c, \, \big\Vert \zeta  \big\Vert_{\infty} \leq 1, \zeta \in \mathbb{R}^s \big\}.
\end{align}
\end{defn} \vspace{6pt}


\begin{defn} \textbf{(Constrained zonotope)} \label{defcz}
Given a generator matrix $G \in \mathbb{R}^{n \times s}$ and a center $c \in \mathbb{R}^n$, a constrained zonotope $\cal{C}=\big<G,c,A_c,b_c\big>$ of dimension $n$  with $s$ generators is represented by \vspace{-3pt}
\begin{align} 
& \cal{C}=\big<G,c,A_c,b_c\big>  \\ \nonumber & \quad  =\big\{x \in \mathbb{R}^n: x=G \, \zeta +c, \, \big\Vert \zeta \big\Vert_{\infty} \leq 1, \,\,\,  {A_c} \zeta=b_c \big\},
\end{align}
where ${A_C} \in \mathbb{R}^{n_c \times s}$, and $b_C \in \mathbb{R}^{n_c}$. 
\end{defn} \vspace{6pt}

\begin{defn} \textbf{(Matrix zonotope)} \cite{ID3}
Given a generator matrix $G \in \mathbb{R}^{n \times ps}$ and a center $C \in \mathbb{R}^{n \times p}$, a matrix zonotope $\cal{M}=\big<G,C\big>$ of dimension $(n,p)$  with $s$ generators is represented by \vspace{-3pt}
\begin{align}
& \cal{M}=\big<G,C\big> \\ \nonumber & \quad \,\, = \big\{X \in \mathbb{R}^{n \times p}:X=\sum\limits_{i=1}^{s} G_i \, \zeta_i +C, \big\Vert \zeta  \big\Vert_{\infty} \leq 1  \big\},
\end{align}
where $G=[G_1,...,G_s]$ and $G_i \in \mathbb{R}^{n \times p}, \,\, i=1,...,s$. 
\end{defn}  \vspace{6pt}

\begin{defn} \textbf{($T$-concatenation of zonotopes)} \cite{ID3}
The concatenation of two zonotopes $\cal{Z}_A$ and $\cal{Z}_B$ is a matrix zonotope $\cal{M}_{AB}$ formed by the horizontal stacking of the two zonopotes. That is, $\cal{M}_{AB}=\big \{[x_A \,\, x_B]: x_A \in \cal{Z}_A, \, x_B \in \cal{Z}_B \big \}$. From this definition, the $T$-concatenation of a zonotope $\cal{Z}_A$ is its concatenation of $\cal{Z}_A$ with itself $T$ times. That is, $\cal{M}_{A^T}=\big \{[x_A^1,..., x_A^T]: x_A^i \in \cal{Z}_A, i=1,..,T \big \}$.
\end{defn}  \vspace{6pt}

\begin{defn} \textbf{(Constrained Matrix zonotope)} \cite{ID3}
Given a generator matrix $G \in \mathbb{R}^{n \times ps}$, a center $C \in \mathbb{R}^{n \times p}$, a constrained matrix zonotope $\cal{K}=\big<G,C,A_C,B_C\big>$ of dimension $(n,p)$  with $s$ generators is represented by \vspace{-3pt}
\begin{align} 
& \cal{K}=\big<G,C,A_C,B_C\big> \\ \nonumber & \quad \,\, = \Big\{X \in \mathbb{R}^{n \times p}:X=\sum\limits_{i=1}^{s} G_i \, \zeta_i +C, \\ \nonumber & \quad \quad \quad \sum\limits_{i=1}^{s} {A_C}_i \zeta_i=B_C, \,\, \big\Vert \zeta  \big\Vert_{\infty} \leq 1  \Big\},
\end{align}
where $G=[G_1,...,G_s], A_C=[{A_C}_1,...,{A_C}_s]$ and $G_i \in \mathbb{R}^{n \times p}, \, {A_C}_i \in \mathbb{R}^{n_c \times p_c} \,\, $, $i=1,...,s$,  
and  $B_C \in \mathbb{R}^{n_c \times p_c}$. 
\end{defn}  \vspace{6pt}


\begin{lem} \label{twoZen} \cite{setcont2}
   For two constrained zonotope $\cal{C}_i=\big<G_i,c_i,A_i,b_i \big>, \, i=1,2$, their Minkowski sum becomes 
   \begin{align} \label{minsum}
       Z \oplus W=\Big<[G_1 \quad G_2],c_1+c_2,\begin{bmatrix}
    A_1 & 0 \\
    0 & A_2
\end{bmatrix},\begin{bmatrix}
    b_1  \\
    b_2
\end{bmatrix} \Big>.
   \end{align}
\end{lem}  \hfill   $\blacksquare$  \vspace{6pt}

\begin{lem} \cite{inclusion} \label{inclusion}
    Consider two constrained zonotopes $\cal{C}_i=\big<G_i,c_i,A_i,b_i \big>$ with $c_i \in \mathbb{R}^{n_i}$, $G_i \in \mathbb{R}^{n_i \times s_i}$, $A_i \in \mathbb{R}^{q_i \times s_i}$ and $b_i \in \mathbb{R}^{q_i}$, $i=1,2$. Then, set inclusion $\cal{C}_1 \subseteq \cal{C}_2$ hold if
    \begin{align} 
        \exists \Gamma \in \mathbb{R}^{s_2 \times s_1}, L \in \mathbb{R}^{s_2}, P \in \mathbb{R}^{q_2 \times q_1}
    \end{align}
such that  \vspace{-12pt}
    \begin{align}
      &  c_2-c_1=G_2 L, \quad G_1=G_2 \Gamma, \nonumber \\ 
      & P A_1=A_2 \Gamma, \quad Pb_1=b_2+A_2 L, \nonumber \\ 
      & |\Gamma|\bar{\textbf{1}}+|L| \le \bar{\textbf{1}}. 
    \end{align}
\end{lem}  \hfill   $\blacksquare$ 

\begin{lem} \label{intmz}
    Consider two constrained matrix zonotopes $\cal{M}^l=\big<G^l,C^l,A^l,B^l \big>$, $G^l \in \mathbb{R}^{n_l \times p_l s_l}$, $C^l \in \mathbb{R}^{n_l \times p_l}$, $A^l \in \mathbb{R}^{n_{ci}  \times p_{cl} s_l}$, and $B^l \in \mathbb{R}^{n_{cl}  \times p_{cl} }$, $l=1,2$. Let $p_{cl}=p_l, \,\, l=1,2$. Then, 
    \begin{align}
        \cal{M}^1 \bigcap \cal{M}^2=\Big< [G^1 \quad 0],C^1,\begin{bmatrix}
            A^1 & 0 \\
            0 & A^2 \\
            G^1 & - G^2
        \end{bmatrix}, \begin{bmatrix}
            B^1 \\
            B^2 \\
            C^2-C^1
        \end{bmatrix}.
    \end{align}
\end{lem}
\noindent \textit{Proof.} The proof is similar to the proof of  \cite{Czen}, which is provided for generalized constrained zonotopes. \hfill   $\blacksquare$

\vspace{-6pt}

\section{Problem Formulation}
Consider the discrete-time system with dynamics 
\begin{equation}\label{system} 
x(t+1) = {A}^{*}x(t) + B^{*} u(t) + w(t),
\end{equation}
where $x(t) \in \cal{S}_s \subset \mathbb{R}^n$ is the system's state, $u(t) \in \mathbb{R}^m$ is the control input, and $w(t) \in  \cal{Z}_w \subset \mathbb{R}^n$ is the additive disturbance. 

The following assumptions are made for the system \eqref{system}. \vspace{3pt}

\begin{assumption}
 $\theta^{*}=[A^{*} \quad B^{*}]$ is uncertain and is only known to belong to a constrained matrix zonotope. That is, $\theta^* \in \cal{M}_{prior}=\big <G_{\theta},C_{\theta},A_{\theta},B_{\theta} \big>$ for some $C_{\theta} \in \mathbb{R}^{n \times (n+m)}$ and $G_{\theta} \in \mathbb{R}^{n \times (n+m)s_{\theta}}$, $A_{\theta} \in \mathbb{R}^{n_{\theta} \times T s_{\theta}}$ and $B_{\theta} \in \mathbb{R}^{n_{\theta} \times T}$. 

\end{assumption}

\vspace{3pt}

\begin{assumption}
    The pair $(A,B)$ is controllable.  
\end{assumption} \vspace{3pt}

\begin{assumption} \label{dist}
   The disturbance set $\cal{Z}_w$ is a constrained zonotope of order $n$ with $s_{w}$ generators. That is, $\cal{Z}_w=\big<G_h,c_h,A_h,b_h\big>$ for some $G_h \in \mathbb{R}^{n \times s_{w}}$, $c_h \in  \mathbb{R}^{n}$, $A_h \in \mathbb{R}^{n_w \times s_w}$, and $b_h \in \mathbb{R}^{n_w}$.
\end{assumption} \vspace{3pt}

\begin{assumption}
  The system's safe set $\cal{S}_s$ is given by either 1) a constrained zonotope
      $\cal{C}_x=\big<G_x,c_x,A_x,b_x \big>$ for some  $c_x \in \mathbb{R}^{n}$, $G_x \in \mathbb{R}^{n \times s_x}$, $A_x \in \mathbb{R}^{q \times s_x}$, and $b_x \in \mathbb{R}^{q}$, or 2) by its equivalent convex polytope $\cal{P}(H_s,h_s)$ for some matrix $H_s \in \mathbb{R}^{q \times n}$ and vector $h_s \in \mathbb{R}^{q}$.  
\end{assumption} \vspace{3pt}

To collect data for learning, a sequence of control inputs, given as follows, is typically applied to the system \eqref{system}
\begin{align} \label{data-u}
U_0 := \begin{bmatrix} u(0) & u(1) & \cdots & u(T-1) \end{bmatrix} \in \mathbb{R}^{m \times T}.
\end{align}
We then arrange the collected $T+1$ samples of the state vectors as \vspace{-9pt}
\begin{align} \label{data-x1}
X := \begin{bmatrix} x(0) & x(1) & \cdots & x(T) \end{bmatrix} \in \mathbb{R}^{n \times (T+1)}.
\end{align}

These collected state samples are then organized as follows 
\begin{align} \label{data-x}
X_0 &:= \begin{bmatrix} x(0) & x(1) & \cdots & x(T-1) \end{bmatrix} \in \mathbb{R}^{n \times T}, \\
X_1 &:= \begin{bmatrix} x(1) & x(2) & \cdots & x(T) \end{bmatrix} \in \mathbb{R}^{n \times T}.  \label{data-z}
\end{align}

Additionally, the sequence of unknown and unmeasurable disturbances is represented as
\begin{equation}\label{data}
W_0 :=  \begin{bmatrix} w(0) & w(1) & \ldots & w(T-1) \end{bmatrix} \in \mathbb{R}^{n \times T},
\end{equation}
which is not available for the control design. We also define
\begin{align} \label{D}
  D_0:= \begin{bmatrix}
X_0 \\ U_0
\end{bmatrix}.
\end{align}
\vspace{-3pt}

\begin{assumption}\label{assumption_5}
The data matrix $X_0$ has full row rank, and the number of samples satisfies $T \geq n+1$.
\end{assumption} \vspace{6pt}

We are now ready to formalize the data-based safe control design problem. The following definition is required. \vspace{6pt}

\begin{defn}
\textbf{Robust Invariant Set (RIS)} \cite{SetB} The set $\cal{P}$ is a RIS for the system \eqref{system} if $x(0) \in \cal{P}$ implies that $x(t) \in \cal{P} \,\,\, \forall t \ge 0, \,\,\,  \forall w \in \cal{Z}_w $.
\end{defn} \vspace{6pt}

\noindent \textbf{Problem 1: Data-based Safe control Design} 
Consider the collected data \eqref{data-u}, \eqref{data-x1} from the system \eqref{system} under Assumptions 1-5. Learn a controller in the form of \vspace{-3pt}
\begin{align} \label{cont}
    u(t)=K x(t),
\end{align}
to ensure that the safe set described in Assumption 4 is an RIS. To this end, use the following two sources of information: 1) prior information (i.e., prior knowledge of the system parameters $\theta^*$ and the disturbance bound $w$), and 2) the posterior information (i.e., collected data).  \vspace{3pt}

To design a controller that guarantees RIS of the safe set (i.e., to ensure that the system's states never leave the safe set), we leverage the concept of $\lambda$-contractive sets defined next. \vspace{3pt}

\begin{defn} \textbf{Contractive Sets} \label{contract} \cite{SetB} Given a $\lambda \in (0,1)$, the set  $\cal{P}$ is $\lambda$-contractive for the system \eqref{system} if $x(t) \in \cal{P} $ implies that $x(t+1) \in \lambda \cal{P}, \,\,\, \forall w \in \cal{Z}_w$.  \vspace{3pt}
\end{defn}

Guaranteeing that a set is $\lambda$-contractive not only guarantees that the set is RIS, but also the convergence of the system's states to the origin with a speed of at least $\lambda$. \vspace{-6pt}

\section{Closed-loop System Representation Using Data and Prior Knowledge}
In this section, we present a novel data-based closed-loop representation of the system that integrates both data and prior knowledge in a systematic manner. 

Since the actual realization of the disturbance sequence $W_0$
is unknown, there generally exist multiple pairs $\theta=[A \quad B]$ which are consistent with the data for 
some disturbance instance $W_0 \in \cal{M}_{\cal{Z}_w^T}$, where
\begin{align} \label{T-dis}
    \cal{M}_{\cal{Z}_w^T}=\big <G_w,C_w,A_w,B_w\big>,
\end{align}
is the constrained matrix zonotope formed by $T$-concatenation of $\cal{Z}_w$, with $G_w \in \mathbb{R}^{n \times T s_w}$, $C_w \in \mathbb{R}^{n \times T}$, $A_w \in \mathbb{R}^{n_{w}  \times T s_w}$, and $B_w \in \mathbb{R}^{n_{w}  \times T }$. 
We denote the set of all such open-loop models $\theta=[A \quad B]$ consistent with both data and prior knowledge by
\begin{align} \label{open-set}
    \Sigma_{X,U_0}=\Big \{\theta \in \cal{M}_{prior}: X_1=\theta D_0+W, \,\, W \in  \cal{M}_{\cal{Z}_w^T} \Big  \}.
\end{align}
For a given control gain $K$, we now define the set of all closed-loop systems consistent with data and prior knowledge as
\begin{align} \label{cl-set}
   \Sigma^K_{X,U_0}= \Big \{A_K: A_K=A+BK, \,\,  [A \quad B] \in \Sigma_{X,U_0} \Big \}.
\end{align}
The following lemma is required for the development of the closed-loop characterization using constrained matrix zonotopes.
\vspace{6pt}

\begin{lem} \label{trans}
    Let $\cal{M}=\big <G,C \big>$ be a matrix zonotope of dimension $(n,p)$ with $s$ generators. Then, the transformation $\cal{M}\circ N =\{XN: X \in \cal{M}\}$ for some matrix $N \in \mathbb{R}^{p \times n}$ is a matrix zonotope of order $(n,n)$ with $s$ generators, defined by  $\cal{M}_N=\big <G \circ N,C N \big>$. \vspace{3pt}
\end{lem}
\noindent \textit{Proof:} The proof is along the lines of \cite{Czen}. For any $X \in \cal{M}$, there exists a $\zeta \in B_{\infty}$ such that $X=\sum\limits_{i=1}^{s} G_i \, \zeta_i +C$. Therefore, $XN=\sum\limits_{i=1}^{s} (G_i N) \, \zeta_i +CN$. By the definition of $\cal{M}_N$, this implies that $XN \in \cal{M}_N$, and since $X$ is arbitrary, $\cal{M} \circ N \subseteq \cal{M}_N$. 
Conversely, for any $Y \in \cal{M}_N$ there exists $\zeta \in B_{\infty}$ such that $Y=G \circ N \, \zeta+CN= \sum\limits_{i=1}^{s} G_i N \, \zeta_i +CN= (\sum\limits_{i=1}^{s} G_i \, \zeta_i +C)N$. Therefore, there exists $X \in \cal{M}$ with $Y=XN$. That is, $Y \in \cal{M} \circ N$, and since $Y$ is arbitrary, $\cal{M_N} \subseteq \cal{M} \circ N$. We conclude that $\cal{M_N}= \cal{M} \circ N$, which completes the proof.  \hfill   $\blacksquare$ \vspace{6pt}

\begin{lem} \label{trans2}
    Let $\cal{K}=\big <G,C,A_C,B_C \big>$ be a constrained matrix zonotope of dimension $(n,p)$. Then,  $\cal{K}\circ N =\{XN: X \in \cal{K}\}$ for some  vector $N \in \mathbb{R}^{p}$ (matrix $N \in \mathbb{R}^{p \times q}$  ) is a constrained zonotope (a constrained matrix zonotope), defined by $\cal{K}_N=\big <G \circ N,CN,\text{Vec}(A_C),\text{Vec}(B_C) \big>$ \big($\cal{K}_N=\big <G \circ N,CN,A_C,B_C \big>$ \big). 
\end{lem}
\noindent \textit{Proof:} The proof is similar to Lemma \ref{trans}. \hfill   $\blacksquare$\vspace{6pt}

Inspired by \cite{Data2}, we parameterize the control gain by a decision variable $G_K$ as $K=U_0 G_K$ where $G_{K} \in \mathbb{R}^{T \times n}$ satisfies $X_0 G_K=I$. That is, we assume $G_K$ satisfies
\begin{align} \label{G}
   D_0 \, G_K=\begin{bmatrix}
I \\ K
\end{bmatrix},
\end{align}
and parametrize the set of all closed-loop systems. \vspace{6pt}

\begin{thm} \label{clrep}
Consider the system \eqref{system}. Let the input-state collected data be given by \eqref{data-u} and \eqref{data-x}-\eqref{data-z}, and the prior knowledge be given by Assumptions 1 and 3. Then, under Assumption 5 and parametrization \eqref{G}, the set \eqref{cl-set} is exactly represented by the following constrained matrix zonotope of dimension $(n,n)$
 \begin{align}\label{clset}
\cal{M}_{cl}=\Big<[G_w \circ G_K \quad 0],(X_1-C_w)G_K,A_C,B_C \Big>,
\end{align}
 with  $s_c=s_{\theta}+s_w$ generators and
\begin{align} \label{AcBc}
  A_C=\begin{bmatrix}
        A_w & 0  \\ 
        0 & A_{\theta} \\
       G_w  & -G_{\theta}\circ D_0
    \end{bmatrix}, \,\, B_c=\begin{bmatrix}
        B_w \\ 
        B_{\theta} \\
        {X}_1-C_{\theta} D_0-C_w
    \end{bmatrix}.  
\end{align}
\end{thm}

\vspace{3pt}
\noindent \textit{Proof: }
Using the control input $u(t)=Kx(t)$ in the system \eqref{system}, the closed-loop system becomes
\begin{align}\label{cl-syst}
 x(t+1)  = (A^*+B^*K) x(t) + w(t).
\end{align}
On the other hand, using the data \eqref{data-x}-\eqref{data-z} and the system \eqref{system}, one has \vspace{-10pt}
\begin{equation}\label{system-data} 
X_1 = {A}^*X_0 + B^*U_0 + W_0.
\end{equation}
Multiplying both sides of this equation by
$G_K$, one has
\begin{align}\label{system-cl1} 
& (X_1-W_0) G_K=(A^* X_0+B^*U_0) G_{K}.
\end{align}
Using $K=U_0 G_K$ and $X_0 G_K=I$, the data-based closed-loop dynamics becomes 
\begin{align}\label{system-cl2} 
A^*+B^*K=(X_1-W_0) G_K.
\end{align}
Since the disturbance realization is unknown, the set of all possible closed-loop systems $A_K=A+BK$ is a subset of the set characterized by $(X_1-W) G_K$ for some $W \in \cal{M}_{\cal{Z}_w^T}$.
However, this leads to a conservative characterization of closed-loop systems, as the set $\cal{M}_{\cal{Z}_w^T}$ is typically conservatively large. To find the exact set $\Sigma^K_{X,U_0}$, we now need to limit this set by the disturbances that can be explained by data and prior knowledge. Using \eqref{G}, we have
\begin{align} \label{sys-cl3}
 A^*+B^*K= \begin{bmatrix}
    A^* & B^*
 \end{bmatrix} \begin{bmatrix}
    I \\ K
 \end{bmatrix}=  \begin{bmatrix}
    A^* & B^*
 \end{bmatrix} D_0 \, G_K.
\end{align}
Comparing \eqref{system-cl2} and \eqref{sys-cl3}, the consistency condition $[A \quad B] \in \Sigma_{X,U_0}$ implies that \vspace{-6pt}
\begin{align} \label{con11}
 \begin{bmatrix}
    A & B
 \end{bmatrix}  D_0 =(X_1-W),
\end{align}
is satisfied for some $W \in \cal{M}_{\cal{Z}_w^T}$ and some $\theta=[A \quad B] \in \cal{M}_{prior}$.  Using \eqref{con11} and Assumption 1, the disturbance set that is consistent with prior knowledge and data becomes
\begin{align}
    {W}={X}_1-\theta D_0 \in \big<G_{\theta}\circ D_0,{X}_1-C_{\theta} D_0,A_{\theta},B_{\theta} \big>=\cal{M}_{d}
\end{align}
where Lemma \ref{trans2} is used to find $\theta D_0$. 
Besides, there might not exist a disturbance solution $\theta$ to the system of linear equation \eqref{con11}. Similar to \cite{ID3}, one can limit the disturbance set 
 $\cal{M}_{\cal{Z}_{{w}}^T}$. We assume the set  $\cal{M}_{\cal{Z}_{{w}}^T}$ is refined based on  \cite{ID3}. Therefore, the disturbance set that is consistent with both Assumptions 1 and 3 and can be explained by data is obtained by ${W} \in \cal{M}_{d}  \bigcap \cal{M}_{\cal{Z}_{{w}}^T}=\cal{M}_{dp}$. Using Lemma \ref{intmz}, it yields
\begin{align} \label{dp}
  &  \cal{M}_{dp}= \Bigg< \big[ G_w  \quad 0 \big], C_w  ,\begin{bmatrix}
        A_w & 0  \\ 
        0 & A_{\theta} \\
       G_w  & -G_{\theta}\circ D_0
    \end{bmatrix}, \nonumber \\ & \quad \quad \begin{bmatrix}
        B_w \\ 
        B_{\theta} \\
        {X}_1-C_{\theta} D_0-C_w 
    \end{bmatrix} \Bigg>=\big<G_{dw},C_{dw},A_{dw},{B}_{dw} \big>.
\end{align}
Therefore, $\Sigma^K_{X,U_0} \subseteq \cal{A}_G$ where
\begin{align}
 &  \cal{A}_G= \Big \{A_K: A_K=(X_1-W) G_K, \,\, W \in \cal{M}_{dp} \Big \}.
\end{align}
Using Lemma \ref{trans2} and \eqref{dp} in the above equation yields \eqref{clset}.

 \hfill   $\blacksquare$ \vspace{6pt}

\begin{rem}
The disturbance set is limited in \cite{ID3} to those that can be explained or generated for some $\theta$, given data. We, however, further limit the set of disturbance to ensure that they can be explained by some $\theta \in \cal{M}_{prior}$ (and not any $\theta$). Set-membership open-loop learning \cite{ID2} can be leveraged to further shrink the constrained zonotope $\cal{M}_{prior}$, and, consequently, the set of closed-loop systems. Starting from a matrix zonotope for the prior knowledge and using set-membership identification, for every data point, the information set is a set of all feasible system models consistent with data and the disturbance bounds, forming a halfspace \cite{ID2}. The constrained zonotope obtained from prior knowledge can then be refined using set-membership identification by finding the intersection of each halfspace with the constrained matrix zonotope, which is another constrained matrix zonotope \cite{setop}. This allows direct and indirect learning integration in a unified framework, and the number of columns of $B_{\theta}$ becomes $T$. \vspace{3pt}
\end{rem} \vspace{3pt}

\begin{rem}
Theorem \ref{clrep} parametrizes the set of all closed-loop systems using a decision variable $G_K$. To check the existence of a solution, by Assumption 5, a right inverse $G_{K}$ exists such that $X_0 \, G_{K}=I$. {Besides, since the rank of $X_0$ is $n$ while at least $n+1$ samples are collected, its right inverse $G_k$  exists and is not unique.} The non-uniqueness of $G_K$ will be leveraged to learn a controller that satisfies safety properties for all closed-loop systems parameterized by $G_K$. 
Another point to note is that learning compact sets of $A$ and $B$ that are consistent with data requires that the data matrix $D_0$ in \eqref{D} be full-row rank, which is stronger than Assumption 5. This is related to the data informativeness in \cite{Data1} in which it is shown that satisfying some system properties directly by learning closed-loop dynamics is less data-hungry than satisfying them indirectly by learning a system model first, which requires satisfying the persistence of excitation \cite{PE1}.  
 \end{rem}

\section{Robust Safe Control Design} 
We design safe controllers for two different cases: 1) when the safe set is represented by a constrained zonotope, and 2) when the safe set is represented by a convex polytope. 

\vspace{6pt}

\begin{cor}
Consider the system \eqref{system} and let the closed-loop system be given by the constrained matrix zonotope \eqref{clset}.    Then, given $x(t) \in \big<G_x,c_x,A_x,b_x\big>$ as the current state, the next state satisfies 
 \begin{align}\label{statezen}
& x(t+1) \in \cal{C}_{cl}=\big<G_{cl},c_{cl},A_{cl},b_{cl} \big>,
\end{align}
where
\begin{align} \label{clzen}
    &G_{cl}= [(G_w \circ G_K) c_x \quad 0 \,\,\,\, (X_1-C_w)G_K G_x \,\,\,\, G_f \,\,\,\, G_h], \nonumber \\ 
   & c_{cl}=(X_1-C_w)G_K c_x+c_h, \nonumber \\ 
   & A_{cl}=\begin{bmatrix}
    \text{Vec}(A_{C}) & 0  & 0 & 0 \\
    0 & A_x & 0 & 0 \\
    0 & 0 & 0 & A_h
\end{bmatrix},  \nonumber \\ 
& b_{cl}=\begin{bmatrix}
    \text{Vec}(B_c) \\ b_x \\ b_h
\end{bmatrix},
\end{align}
where $G_{f}$ is formed using $G_w \circ G_k$ and $G_x$ and $A_C$, $B_c$, $A_x$ and $b_x$, and $A_C$ and $B_C$ are defined in \eqref{AcBc}.
\end{cor}
\noindent \textit{Proof:}
Using \eqref{clset}, Assumption 3, and $x(t) \in \big<G_x,c_x,A_x,b_x\big>$ in the system dynamics \eqref{system}, one has
\begin{align}
    x(t+1) \in \cal{M}_{cl} \times \cal{C}_x \oplus \cal{Z}_w.
\end{align}
Using proposition 2 in \cite{ID3} one has
\begin{align}
   & \cal{M}_{cl} \times \cal{C}_x \subseteq \cal{C}_{clw}=\Big<[(G_w \circ G_K) c_x  \quad 0
   \nonumber \\ & \quad  \quad \quad \quad \quad \quad \quad \quad \quad (X_1-C_w)G_K G_x \quad G_f], \nonumber \\ &  (X_1-C_w)G_K c_x,  \begin{bmatrix}
    \text{Vec}(A_{C}) & 0  & 0 \\
    0 & A_x & 0
\end{bmatrix},\begin{bmatrix}
   \text{Vec}(B_c) \\ b_x
\end{bmatrix} \Big> \nonumber \\  &
=\big<G_{clw},c_{clw},A_{clw},b_{clw} \big>.
\end{align}
where $G_f$ is formed similar to Proposition 2 of \cite{ID3}. Therefore, $x(t+1) \in \cal{C}_{clw} \oplus \cal{Z}_w$. Using Lemma \ref{twoZen}, for the constrained zonotope $\cal{Z}_w=\big<G_h,c_h,A_h,b_h \big>$ and the constrained zonotope $\cal{C}_{clw}=\big<G_{clw},c_{clw},A_{clw},b_{clw} \big>$, one has $\cal{C}_{clw} \oplus \cal{Z}_w =\Big<[G_{clw} \quad G_h],c_{clw}+c_h,\begin{bmatrix}
    A_{clw} & 0 \\
    0 & A_h
\end{bmatrix},\begin{bmatrix}
    b_{clw}  \\
    b_h
\end{bmatrix} \Big>=\cal{C}_{cl}$, which completes the proof. \hfill   $\blacksquare$ 
 \vspace{2pt}

\begin{thm}
Consider the system \eqref{system} under Assumptions 1-5 with the safe set  $\cal{C}_x=\big<G_{x},c_{x},A_{x},b_{x}. \big>$ Then, Problem 1 is solved using the controller $u(t)=U_0 G_K x(t)$ if there exist $G_K$, $\Gamma$, $L$ and $P$ satisfying \vspace{-2pt}
\begin{align}
   & (I-(X_1-C_w)G_K) c_x-c_h=\lambda G_x L, \nonumber \\
   & G_{cl}=\lambda G_x \Gamma, \nonumber \\
    & P A_{cl}=A_x \Gamma, \nonumber \\
    & P b_{cl}=\lambda b_x+A_x L, \nonumber \\
    &|\Gamma|\bar{\textbf{1}}+|L| \le \bar{\textbf{1}}, \nonumber \\
&
X_0 G_K=I.
\end{align}
where $G_{c}$, $c_{cl}$, $A_{cl}$, $b_{cl}$ are defined in \eqref{clzen}. 
\end{thm}

\noindent \textit{Proof:}  Based on definition of the $\lambda$-contractivity in Definition \ref{contract} is satisfied if $\cal{C}_{cl}$ defined in \eqref{statezen} is contained in $\lambda$-scaled level set of the safe set $\cal{C}_x=\big<G_x,c_x,A_x,b_x \big>$. The $\lambda$-scaled level set of the constrained zonotope is obtained by scaling its boundaries. The safe set is $\cal{C}_x=\{x(t):x(t)=c_x+G_x \zeta,  \,\, \|\zeta\| \le 1, A_x \zeta =b_x \}$. $\lambda$-scaled level set of the safe set is obtained by scaling the boundaries of set, which gives $\{x(t):x(t)=c_x+G_x \zeta,  \,\, \|\zeta\| \le \lambda, A_x \zeta = \lambda b_x \}$, or equivalently, a constrained zonotope $\cal{C}_{x\lambda}=\big<\lambda G_x,c_x,A_x, \lambda b_x$ \big>. Therefore, the proof boils down to ensuring $\cal{C}_{cl} \subseteq \cal{C}_{x\lambda}$. The rest of the proof follows Lemma \ref{inclusion}.  \hfill   $\blacksquare$ \vspace{6pt}

\begin{thm}
Consider the system \eqref{system} under Assumptions 1-5 and assume that the safe set is represented by the convex polytope $\cal{P}(H_s,h_s)$. Then, Problem 1 is solved using the controller $u(t)=U_0 G_K x(t)$ if there exist $G_K$, $\rho$  and $P$ satisfying \vspace{-13pt}
\begin{align} \label{LPf}
&  \min\limits_{P,G_{K},\rho} \,\,\, \rho \nonumber \\
&{P} h_s \le \lambda  h_s-H_s c_h-\rho \, l -y, \nonumber  \\
&{P} H_s  = H_s (X_1-C_w) G_K, \nonumber \\
& \big\Vert G_K \big\Vert  \le \rho, \nonumber \\
&
X_0 G_K=I,
\nonumber \\
& P \ge 0,
\end{align} 
where $y=[y_1,...,y_q]^T$ and $l=[l_1,...,l_q]^T$, and $y_j$ and $l_j$ are defined as \vspace{-6pt}
\begin{align} \label{lj}
& y_j  = \sum\limits_{i=1}^{s_w} | {H_s}_j {G_h}_i|, \nonumber \\
 & l_j  = \max\limits_{\beta}  \Big(\sum\limits_{i=1}^{s_c}  \Big[\big| h_{sj} \big|  \big\Vert  G_{wi} \big\Vert \quad 0 \Big]    \,  \beta_i  \Big) \nonumber \\ &  
 \quad \quad {\rm{s}}{\rm{.t}}{\rm{.}} \,\, \sum\limits_{i=1}^{s_c} \bar{A}_{c_i}\beta_i=\bar{b}_{c}, \nonumber \\  &  \quad \quad \quad \quad  \big\Vert \beta  \big\Vert_{\infty} \leq 1, \,\,\, \beta_i \ge 0, i=1,...,s_c
\end{align}
where 
\begin{align} \label{acbc}
   \bar{A}_c= \begin{bmatrix}
    \text{Vec}(A_C) & 0 \\ 0 & A_h
\end{bmatrix}, \quad \bar{b}_c= \begin{bmatrix}
    \text{Vec}(B_c) \\ b_h
\end{bmatrix},
\end{align}
and $A_C$ and $B_C$ defined in \eqref{AcBc}.
\end{thm}
\noindent \textit{Proof:}
Using \eqref{clset}, Lemma \ref{trans2} and Lemma \ref{twoZen}, one has 
\begin{align} \label{next2}
 &   x(t+1) \in \Big<[{G}_w \circ G_K \circ x(t) \quad 0 \quad G_h], \nonumber \\ & \quad \quad \quad  \quad \quad (X_1-C_w)G_K x(t)+c_h,\bar{A}_c,\bar{b}_c\Big>.
\end{align}

 By definition, $\lambda$-contractivity is satisfied if $\gamma_j=\max\limits_{x(t)} \, H_{sj} x(t+1) \le \lambda {h_c}_j, \,\, j=1,...,q$, whenever $H_s x(t) \le h_s$. Using \eqref{next2}, define
\begin{align} \label{bargam}
 & \bar \gamma_j  = \max\limits_{\beta} \max\limits_{x(t)}  \Big( {H_s}_j (X_1-C_w)G_K \, x(t)+{H_s}_j c_h+ \nonumber \\ &  \sum\limits_{i=1}^{s_c}  \Big [\big| {H_s}_j G_{wi} G_K x(t) \big| \quad 0 \Big] \beta_i \big| +  \sum\limits_{i=1}^{s_w}  |{H_s}_j  {G_h}_i | \Big) \nonumber \\  &
 {\rm{s}}{\rm{.t}}{\rm{.}} \,\, \sum\limits_{i=1}^{s_c} \bar{A}_{c_i} \beta_i=\bar{b}_{c},  \quad \big\Vert \beta  \big\Vert_{\infty} \leq 1, \nonumber \\  & \quad \quad H_s x(t) \le h_s,
\end{align}
where the last two terms of the cost function are obtained using the fact that for $\zeta=[\beta^T \,\,\, \eta^T] \in \mathbb{R}^{s_c+s_w}$, one has
\begin{align} \label{ineql}
   & {H_s}_j \, [G_w \circ G_K \circ x(t) \quad 0  \quad G_h] \zeta  \le  \nonumber \\ & 
    \sum\limits_{i=1}^{s_{c}}  \Big[ |{H_s}_j G_{wi} G_K x(t)| \quad 0  \Big] \,  |\beta_i|  +  \sum\limits_{i=1}^{s_{w}} | {H_s}_j  {G_h}_i |.
\end{align}
Based on \eqref{ineql}, $\gamma_j \le \bar \gamma_j$, and, thus, Problem 1 is solved if $\bar \gamma_j \le \lambda h_{cj}, \, j=1,...,q$. To find a bound for the term depending on $\beta_i$ in the cost function, since ${H_s}_j x(t) \leq {h_s}_j$, one has $\|x(t)\| \leq \frac{|{h_s}_j|}{\|{H_s}_j\|}$. Using this inequality, one has $\big| {H_s}_j G_{wi} G_K x(t)\big| \le \big| h_{sj} \big|  \big\Vert { {G_w}_i}  \big\Vert \big\Vert {G_K}  \big\Vert$. Therefore, $\bar \gamma_j \le \hat \gamma_j$ where
\begin{align} \label{hatgamma}
 & \hat \gamma_j  =  \max\limits_{x(t)}  \Big( {H_s}_j (X_1-G_w)G_K \, x(t)+{H_s}_j c_h+  \big\Vert {G_K}  \big\Vert l_j+y_j \Big) \nonumber \\  & 
 {\rm{s}}{\rm{.t}}{\rm{.}}   \quad H_s x(t) \le h_s,
\end{align}
where $l_j$ and $y_j$ are defined in \eqref{lj}, inside of which the absolute value of $\beta_i$ in \eqref{bargam} is removed by adding the constraint $\beta_i\ge 0$. 
Using duality, $\lambda$-contactivity is satisfied if $\tilde \gamma_j \le \lambda {h_c}_j$ where \vspace{-8pt}
\begin{align} 
&\tilde \gamma_j  = \min\limits_{\alpha_j} \alpha_j^T \, h_s+  {H_s}_j c_h+ l_j \big\Vert {G_K} \big\Vert+y_j, \label{a1} \nonumber \\&
{\alpha_j}^T H_s = {H_s}_j (X_1-C_w) G_K,  \nonumber \\&
{\alpha_j}^T \ge 0,  
\end{align}
where $\alpha_i \in \mathbb{R}^q$. Define $
{P}=[\alpha_1,....,\alpha_q]^T \in \mathbb{R}^{q \times q}$. $P$ is non–negative since $\alpha_i$ is non-negative for all $i=1,...,q$. Therefore, using the dual optimization,  $\lambda$-contractivity is satisfied if \eqref{LPf} is satisfied. \hfill   $\blacksquare$ \vspace{-6pt}

\section{Simulation Results}
Consider the following discrete-time system used in \cite{example}
\begin{align}
& x_1(t+1) = a_1^* x_1(t) + a_2^* x_2(t)+b_1^* u(t)+ w(t),  \nonumber \\ &
x_2(t+1) = a_3^* x_1(t) +a_4^* x_2(t)+b_2^* u(t).
\end{align}
where the actual but unknown values of the system are $a_1^*=0.8, \,\, a_2^*=0.5, \,\, a_3^*=-0.4, \,\, a_4=1.2 \,\, b_1^*=0, \,\, b_2^*=1$.
The safe set in \cite{example} is considered as a polyhedral set $\cal{C}_x=\{x: H_sx \le h_s \}$ where $h_s=[1,1,1,1]^T$ and $H_s$ is a $4 \times 2$ matrix defined in \cite{example}.
 Since the safe set is symmetric here, it is equivalent to a zonotope $\cal{C}_x=\big<G_x,c_x,0,0\big>$. The contractility level is chosen as $\lambda=0.95$. 
It is also assumed that the prior knowledge of the system parameters provided the bounds $a_1^* \in [0.6 \quad 1], \quad a_2^* \in [0.35 \quad 0.65], \quad a_3^* \in [-0.3 \quad -0.5], \quad a_4^* \in [1 \quad 1.4], \quad b_1^* \in [-0.1 \quad 0.1], \,\,  \text{and} \,\,\, b_2^* \in [0.8 \quad 1.2]$. A matrix zonotope can be found for this prior knowledge since the prior knowledge is provided as box constraints.
A controller in the form of $u(t)=K [x_1 \,\, x_2]$ is then learned using Theorem 3. We compared the results for the case where the prior knowledge is used and the case where it is ignored. The comparison is performed in terms of the disturbance level they can tolerate and the speed of convergence (i.e., $\lambda$) they can achieve under the same level of disturbance. To this end, the unknown disturbance set is assumed $\cal{Z}_w =[-b, b]^2$, with $b$ as the disturbance level. A zonotope with a symmetric positive definite generator is formed for this disturbance set. Our simulation results showed that while the disturbance bound that the learning algorithm can tolerate and provide a solution for the case where the prior knowledge is ignored is $b=0.05$ for a fixed $\lambda=0.98$, the disturbance level that the control design algorithm that accounts for prior knowledge can tolerate increase to $b=0.08$ for the same speed level $\lambda=0.98$. Besides, for a disturbance level $b=0.04$, while the minimum value of $\lambda$ that can be achieved is $0.89$ for the case where no prior knowledge is used, this value decreases to $0.76$ when prior knowledge is leveraged. Figure 1 shows how using prior knowledge can improve performance by finding a lower value $\lambda$ for which the optimization is feasible. The disturbance level is considered as $b=0.03$ in this case.

\begin{figure}[t!]
    \vspace{-10pt}
        \centering
        \subfloat[]{{\includegraphics[width=\linewidth,height=2in, trim=0cm 6.0cm 0cm 4.2cm, clip]{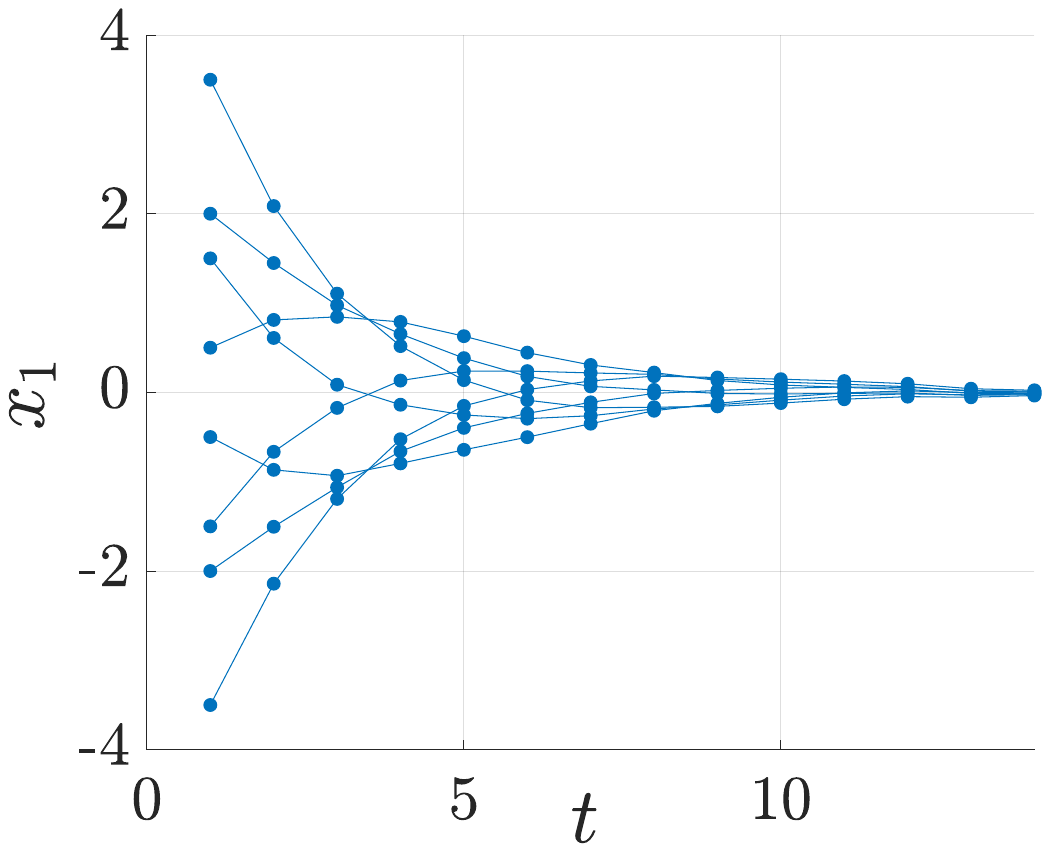} }}%
\vspace{-30pt}
        \centering
        \subfloat[]{{\includegraphics[width=\linewidth,height=2 in, trim=0cm 6.0cm 0cm 4.2cm, clip]{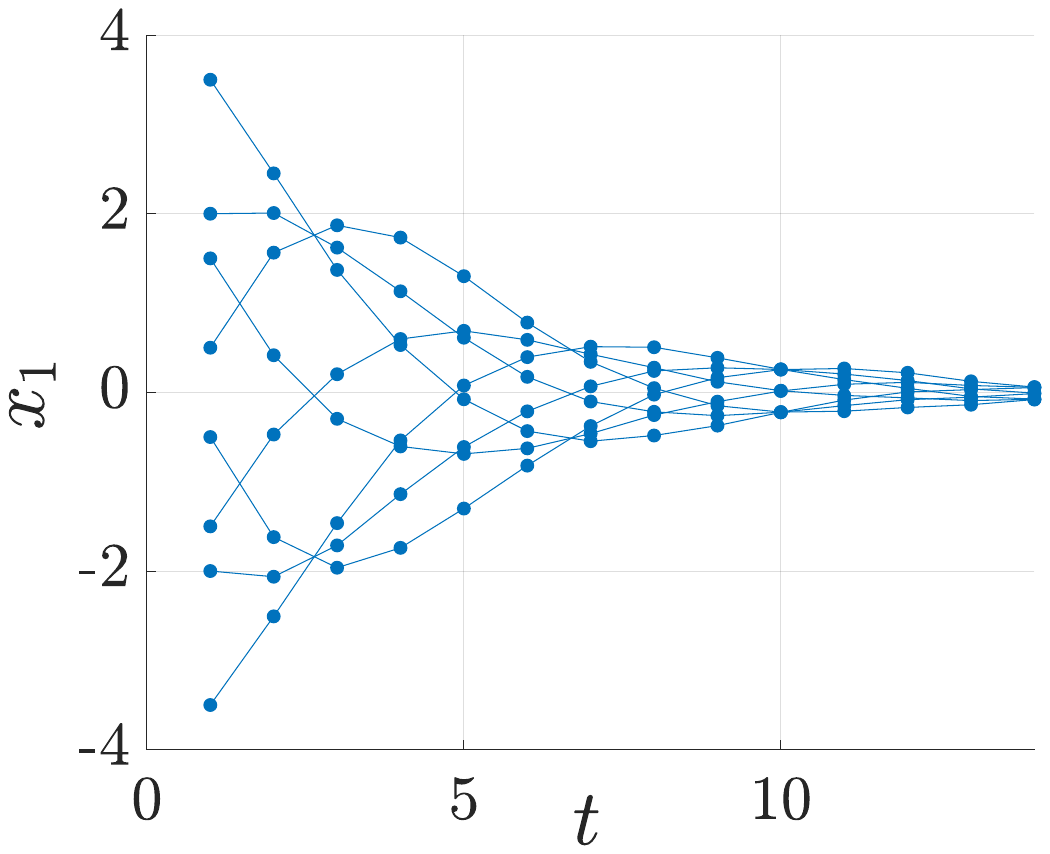} }}%
    \caption{Comparison of the performance of the case where prior knowledge is used (Figure (a)) with the case where prior knowledge is ignored (Figure (b)) for the same level of disturbance.}
    \label{fig:alpha_results}
\end{figure}

\section{conclusion} A novel approach is presented to integrate prior knowledge and open-loop learning with closed-loop learning for safe control design of linear systems. Assuming that the disturbance belongs to a zonotope and without accounting for prior knowledge, we show that the set of the closed-loop representation of systems can be characterized by a matrix zonotope. We then show how to add prior knowledge into this closed-loop representation by turning the matrix zonotope into a constrained matrix zonotope. The equality constraints added by incorporating prior knowledge limit the set of closed-loop systems to those that can be explained by prior knowledge, therefore reducing its conservatism. We then leveraged a set inclusion approach to impose constrained zonotope safety. 

\bibliographystyle{IEEEtran}

\end{document}